\documentclass[amssymb,aps,prd,nobibnote,twocolumn,preprintnumbers]{revtex4}

 \usepackage{here}

\usepackage[dvipdfmx]{graphicx}
\usepackage{amssymb,amsfonts,amsmath,bm,color,multirow,cases,empheq,hyperref}
\usepackage{mathrsfs}

\usepackage{enumerate}
\usepackage{ulem}
\usepackage{afterpage}

\hypersetup{%
 setpagesize=false,
 bookmarksnumbered=true,%
 bookmarksopen=true,%
 colorlinks=true,%
 linkcolor=blue,%
 citecolor=red}

\newcommand{\fr}[1]{\frac{1}{#1}}

\newcommand{\nonum}{\nonumber\\ }

\newcommand{\cout}[1]{}

\newcommand{\arrayL}[1]{\left(\begin{array}{#1}}
\newcommand{\arrayR}{\end{array}\right)}
\newcommand{\arrayLb}[1]{\left[\begin{array}{#1}}
\newcommand{\arrayRb}{\end{array}\right]}

\begin{document}

\title{
 A Capped Black Hole in Five Dimensions}
\author{Ryotaku Suzuki}
\email{sryotaku@toyota-ti.ac.jp}
\author{Shinya Tomizawa}
\email{tomizawa@toyota-ti.ac.jp}
\affiliation{\vspace{3mm}Mathematical Physics Laboratory, Toyota Technological Institute\vspace{2mm}\\Hisakata 2-12-1, Tempaku-ku, Nagoya, Japan 468-8511\vspace{3mm}}

\begin{abstract}

We present the first non-BPS exact solution of an asymptotically flat, stationary spherical black hole having  domain of outer communication with nontrivial topology
 in five-dimensional minimal supergravity. It describes a charged rotating black hole capped by a disc-shaped bubble.
The existence of the ``capped black hole'' shows
the non-uniqueness of spherical black holes.

\end{abstract}

\date{\today}
\preprint{TTI-MATHPHYS-24}
\maketitle

According to the uniqueness theorem for charged rotating black holes in the bosonic sector of five-dimensional minimal supergravity~\cite{Tomizawa:2009ua}, under the assumptions of the existence of
two commuting axial isometries and spherical topology of horizon cross-sections, an asymptotically flat, stationary charged rotating black hole with a non-extremal horizon is uniquely characterized by the mass, charge, and two independent angular momenta and therefore is described by the five-dimensional Cveti\v{c}-Youm solution~\cite{Cvetic:1996xz}. 
Therefore, it seems that there is no room for another spherical black hole in a class of asymptotically flat,  regular solutions with the same symmetry.
However, the topological censorship theorem~\cite{Friedman:1993ty} gives us the possibility of another black hole with spherical topology, since in the uniqueness theorem~\cite{Tomizawa:2009ua}, the intersection of the exterior region of a black hole and the timeslice $\Sigma$ is assumed to have the trivial topology ${\mathbb R}^4\setminus {\mathbb B}^4$.

The topological censorship theorem says that the domain of outer communication (DOC) must be simply connected. 
This implies that for four dimensions, the topology of the black hole exterior region admits a trivial topology of ${\rm DOC}\cap \Sigma={\mathbb R}^3\setminus {\mathbb B}^3$, 
whereas for higher dimensions, the DOC can have nontrivial topologies in the sense that timeslice can have homology groups of a higher rank. 
Even in higher dimensions, for static asymptotically flat spacetimes, a black hole solution with such nontrivial DOC does not exist since the uniqueness theorems~\cite{Gibbons:2002bh,Gibbons:2002av} say that the Schwarzschild-Tangherlini solution~\cite{Tangherlini:1963bw} and the higher-dimensional Reissner-Nordstr\"om solution is the only vacuum black hole solution and the charged black hole solution, respectively.  Therefore, if such a solution exists, it must not  lie in a class of static but stationary solutions.
In fact, Kunduri and Lucietti~\cite{Kunduri:2014iga} have constructed four parameter family of a supersymmetric black hole solution
such that it has a spherical horizon topology and two bubbles of a 2-cycle in the DOC in the five-dimensional minimal supergravity.

In this letter, we show the existence of an exact solution of an asymptotically flat, stationary, non-supersymmetric black hole with the horizon cross section of trivial topology $S^3$ and DOC of nontrivial topology 
 in the bosonic sector of the five-dimensional minimal supergravity. 
As shown below, the horizon looks like being capped by a disc-shaped bubble~\footnote{A set of fixed points of a $U(1)$ isometry that is not a point but has a finite size is often referred to as a bubble.},
and so we call the black hole a ``capped black hole''.
Our solution is regular in the sense that it does not have
any one of curvature singularities, conical singularities, Dirac-Misner string singularities, or orbifold singularities on and outside the horizon, and, in addition, is free from closed timelike curves (CTCs). 
This is the first example of such a black hole as a non-Bogomol'nyi-Prasad-Sommerfield (BPS) solution.

The action of the bosonic sector in the five-dimensional minimal supergravity is given by
\begin{align}
\hspace{-0.5cm}S =\frac{1}{16 \pi G_5} \int \left[ R\star 1-\frac{1}{2}F \wedge \star F -\frac{1}{3\sqrt{3}} F\wedge F\wedge A\right],\hspace{-0.5cm}
\end{align}
where $F:=dA$.
Applying the Harrison transformation of eq.~(119) in Ref.~\cite{Bouchareb:2007ax} to the vacuum rotating black lens
constructed in Ref.~\cite{Chen:2008fa}~(\footnote{As written in the footnote 4 of ref.~\cite{Chen:2008fa},  Dirac-Misner string singularities are removed in the solution~(3.1) in a certain parameter choice. However, we have removed them  not before  but after the Harrison transformation
in terms of one timelike vector $\partial_t$ and one spacelike Killing vector $\partial_\psi$.}),
 we can obtain the metric and gauge potential for a capped black hole, respectively, as~\footnote{See Supplemental Material at [URL] to use the solution data in the Mathematica format.}
\begin{align}
 ds^2 =& -D^{-2} H(y,x)  H(x,y)^{-1}(dt + \Omega'_\psi d\psi + \Omega'_\phi d\phi)^2\notag\\
  &\hspace{-1cm}+D H(y,x)^{-1}\left[ F(y,x)d\psi^2 - 2J(x,y)d\psi d\phi -F(x,y)d\phi^2\right]\notag\\
 &\hspace{-1cm}
 + \frac{\ell^2 D H(x,y)}{4(1-\gamma)^3(1-\nu)^2(1-a^2)(x-y)^2}\left[ \frac{dx^2}{G(x)}-\frac{dy^2}{G(y)}\right],\hspace{-0.5cm}
\end{align}
\begin{align}
\hspace{-0.5cm} A =& \sqrt{3}cs D^{-1} H(x,y)^{-1}\left[ \left\{H(x,y)-H(y,x)\right\}dt \right.\notag\\
 &\left. \hspace{-0.3cm} - \left\{c H(y,x) \Omega_\psi(x,y)-s H(x,y) \Omega_\phi(y,x)\right\}d\psi \right.&\notag\\
& \left. \hspace{-0.3cm} - \left\{ c H(y,x)\Omega_\phi(x,y)-s H(x,y)\Omega_\psi(y,x)\right\}d\phi \right],
\end{align}
where  $(c,s) := (\cosh \alpha,\sinh \alpha)$ with the parameter $\alpha$ corresponding to the electric charge. 
The functions $D$ and  $(\Omega'_\psi,\Omega'_\phi)$ are given by 
\begin{align}
&D=(c^2 H(x,y)  - s^2 H(y,x))/H(x,y),\\
&\Omega_\psi' = c^3 \Omega_\psi(x,y) -s^3 \Omega_\phi(y,x),\\
&\Omega_\phi' = c^3 \Omega_\phi(x,y)-s^3 \Omega_\psi(y,x).
\end{align}
Remaining functions $(G,H,F,J,\Omega_\phi,\Omega_\psi)$ are written as
\begin{flalign}
G(\xi)& = (1-\xi^2)(1+\nu \xi),
\end{flalign}
and
\begin{widetext}
\begin{flalign}
H(x,y)&=\left[ \nu  b^2 c_1^2  (1+\nu )^2(\gamma -\nu ) (1-\gamma )  (1-x^2)
+ \nu(1-\gamma )   (\gamma -\nu ) \left(b (1-\nu )^2 (1-x)-2 c_1 (1+\nu  x)\right)^2\right.\nonum
& \hspace{-0.5cm}+ b^2 c_1^2 (1+\nu )^3 (x+1) (\gamma -\nu )^2 \bigr](1+y)^2
-\bigl[d_5  (1-x)^2+d_6(1-x^2)+ d_7(1+x) (1+\nu  x) \bigr](1+y)\nonum
& \hspace{-0.5cm}+ 4 \left(1-a^2\right) (1-\gamma )^3 (1-\nu )^4 (1-x)
 +2c_2^2  (1-\gamma )  (1-\nu )^2   \left(1-x^2\right)
-4d_1  (1-\gamma ) (1-\nu)^2 (1+\nu ) (1+x),\label{eq:H2}
\\
F(x,y) &=2 \ell^2 (1-a^2)^{-1}(x-y)^{-2} \left[
4 \left\{\left(1-a^2\right)^2 (y-1) (1-\gamma )^3 (1-\nu )^3-(1+y) d
   _1^2\right\}(1+y \nu ) G(x)\right.\nonum
&+\nu^{-1}(1-\gamma)^{-1}G(x)G(y)\left\{\left(1-a^2\right) (1-\gamma )
   d _4-\nu  c _2^2(a-b)^2  (1-\gamma )^2 (\gamma -\nu )y +\nu x (\gamma -\nu )   \left(c _1 c _3 - b d _1\right){}^2\right\}\nonum   
&+4\left\{(1-\nu ) c _2-(1-a b) (\gamma -\nu ) (1+\nu ) c _1\right\}^2 (1+x \nu )  (1+x)G(y)\nonum
&\left.+ \nu^{-1} (1-\nu )^3 (\gamma -\nu ) \left\{d _3^2(1-x^2)G(y) - c _3^2(1-y^2)G(x) \right\}\right],\\
J(x,y) &=  2\ell^2(1+x)(1+y)(1-a^2)^{-1}(x-y)^{-1}
\left[4  d _1 
   \left\{(a-b) (1-\gamma ) (\gamma -\nu ) (1+\nu )-a d _2\right\}(1+ \nu x) (1+ \nu y)\right.\nonum
&\left. -d _3 c _3 (1-\nu
   )^3 (\gamma -\nu ) (1-x) (1-y)-c _2(a-b) (\gamma -\nu )   \left(c _1 c _3-b d
   _1\right)(1-x) (1-y) (1+\nu x  ) (1+\nu y)\right],\\
\Omega_\psi(x,y)& = v_0 \ell (1+y)(1-\nu)\nu^{-1} H(y,x)^{-1}\left[
c _2 \left(c _1 c_3-b d _1\right)(1-x) (1+x \nu ) (1+y \nu ) \right.\nonum
&\left.- d _2 c _3(1-\nu )^2  (1-x)+d _1(1+x \nu )  \left\{2 \nu(1-a b)  (1-\gamma ) (1+\nu )(1+x)+c _3(1-3   \nu -x (1+\nu )) \right\}
\right],\\
 \Omega_\phi(x,y) & = v_0 \ell (1+x)H(y,x)^{-1}\left[ d _1 b (1+x) \left\{d _2(1+y) (1+y \nu ) +\nu  c _3\left(1-y^2\right) (1-\nu ) \right\}\right.\nonum
&\left.+2 (a-b)(1+\nu)^{-1} (1-\gamma)^2 \left\{2 d _1(1+x \nu )   (1+y \nu )^2 
-\nu c _3(1-\nu )^2 (1-y)  (x+y+\nu +x y \nu ) \right\}\right],
\end{flalign}
where the constants $v_0,c_p,d_q\ (p=1,2,3,q=1,\ldots,7)$ are given by
\begin{align}
&v_0 := \sqrt{2(\gamma^2-\nu^2)(1-a^2)^{-1}(1-\gamma)^{-1}},\quad
c _1 := a(1-\gamma)+b(\gamma - \nu)   ,\quad 
c _2 := 2 a\nu(1-\gamma )  +b (\gamma -\nu ) (1+\nu ),\nonum
&c _3 := 2  \nu  (1-\gamma ) +b^2 (\gamma -\nu ) (1+\nu ),\quad
d _1 := c _1^2(\nu +1) -(1-\gamma ) (1-\nu )^2, \quad
d _2:= bc _1 (\nu +1)  (\gamma -\nu )+2\nu (1-\gamma )   (1-\nu )  ,\nonum
&  d _3 := b(1-a^2)  (1-\gamma ) (\nu +1)-a c _3,\quad
d _4 := b^2 (\gamma -\nu )   \left[c _1^2(\nu +1)^2  \left(1-\nu^2-3 (1-\gamma ) \nu \right)-(1-\gamma ) (1-\nu )^4 (2 \nu +1)\right]\nonum
&+(1-\gamma   ) \left[\left(c _2(1-\nu ) -2 \nu ^2 c _1\right){}^2-4 \nu ^2 c _1^2 \left(3 \nu^2+1-\gamma  (\nu +2)\right)\right],\quad
\nonum
&d_5 := (1-\gamma ) (1-\nu )^3 [(\gamma -3 \nu ) \left(b^2 (\nu -1) (\gamma -\nu )-c_1^2\right)
-2 b c_1 (3 \nu -1) (\gamma -\nu )],\nonum
&d_6/(1-\nu):=d_7/2  := c_1(1-\gamma ) \left(1-\nu ^2\right) \left(c_2-(1-\gamma ) (a-b) (\gamma -\nu )\right).
\end{align}

\end{widetext}

\begin{figure}[t]
\includegraphics[width=6.2cm]{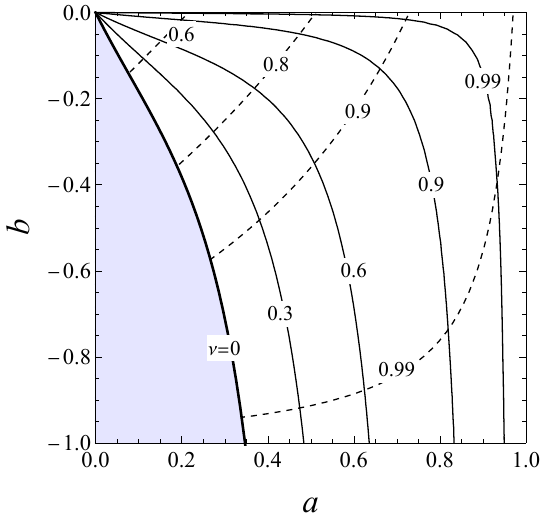}
\caption{
The allowed region for $(a,b)$ in eq.~(\ref{eq:a,b}). The thick and dashed curves correspond to $\nu = {\rm constant}$ and $\tanh\alpha\ = {\rm constant}$, which are given by eqs.~(\ref{eq:Bd-sol}) and (\ref{eq:nodms}), respectively. 
\label{fig:paramregion}}
\end{figure}

The coordinates $(x,y,t,\psi,\phi)$ have the ranges $-1\leq x \leq 1,-1/\nu \leq y \leq -1$,  $-\infty<t<\infty$, $0\le \psi,\phi\le 2\pi$ with the asymptotic infinity at
$x\to y\to -1$ and the event horizon at  $y=-1/\nu$. 
Moreover, 
$x=-1$ and $y=-1$ correspond to the rotation axes around the $\phi$ direction and the $\psi$ direction, respectively, which extend to the asymptotic infinity, and $x=1$ corresponds to the disc-shaped bubble touching the horizon at $(x,y)=(1,-1/\nu)$. 
The parameters $(\ell,\nu,\gamma,a,b,\alpha)$ have the ranges $\ell>0, \ 0<\nu<\gamma<1$, $-\infty<\alpha<\infty$ and 
\begin{align}
\left\{ \begin{array}{ll} 
\displaystyle \frac{a(a-2)(1-a+a^2)}{(1-2a)(1+2a-2a^2)}<b<0 & (0<a<a_*)\\
-1<b<0 &(a_* \leq a <1) 
\end{array},\right.
\label{eq:a,b}
\end{align}
with the constraints
\begin{align}
   \tanh^3 \alpha = (a-b)/(1-ab),\label{eq:nodms}
\end{align}
\begin{align}
\begin{split}
&\nu = 1- \frac{2b(1-a^2)^2}{b\left(2a^2 (a-1)^2+1\right) +a\left((a-1)^3-1\right) },\\
&\gamma =\nu+\frac{(1-\nu)  \left(1-a+a^2 \right)}{1-(1+2b)a+(1+b)a^2},
\end{split}
\label{eq:Bd-sol}
\end{align}
where
$a_*=0.347\ldots$ is a root of $a^3-3a+1=0$.
Thus, the conditions~(\ref{eq:nodms}) and (\ref{eq:Bd-sol}) reduce the number of the independent parameters from six to three. 
As seen below, $\ell$ denotes the scale merely, and $\nu$ controls the size of the horizon or the smallness of the bubble. 
$\alpha$ is related to an electric charge via below Eq.~(\ref{eq:MQ}).
Figure~\ref{fig:paramregion} illustrates the parameter region in the $(a,b)$-plane.

The asymptotic infinity lies at $x\to y\to-1$ since in terms of the standard spherical coordinates $(r,\theta)$ defined as $(x+1,y+1)=\frac{4(1-\nu)\ell^2}{r^2}(\cos^2\theta,\sin^2\theta)$, the metric asymptotes to the five-dimensional Minkowski metric  at $r\to \infty\ (x\to y\to -1)$  as $ds^2\simeq -dt^2+dr^2+r^2(d\theta^2+\cos^2\theta d\phi^2+\sin^2\theta d\psi^2)$.
The metric has the apparent singularities at $x=\pm1,y=-1$, but these are merely coordinate singularities. 
In particular, the Killing vectors $\partial_\phi$ and $\partial_\psi$ vanishes at $x=-1$ and $y=-1$, respectively, which correspond to the rotation axes around the $\phi$ direction and the $\psi$ direction, respectively, and extend to asymptotic infinity. 
The choice of the $2\pi$ periodicities for $\phi,\psi$ guarantees the absence of conical singularities on these rotation axes.

Under the conditions~(\ref{eq:nodms}) and (\ref{eq:Bd-sol}), the Killing vector
$\partial_{\phi'}:=\partial_\phi + \partial_\psi$ vanishes at $x=1\ (-1/\nu\le y\le -1)$, and hence there do not exist Dirac-Misner string singularities.
Furthermore, from eqs.~(\ref{eq:Bd-sol}), we can find $\Delta \phi'=2\pi$, which means the absence of conical singularities at $x=1$.
At $x=1$, another Killing vector $\partial_\psi$ vanishes at $(x,y)=(1,-1)$ but does not vanish for $-1/\nu \le y<-1$ including $(x,y)=(1,-1/\nu)$, and 
then since $x=1\ (-1/\nu\le y\le -1)$ where the $U(1)$ Killing vector $\partial_{\phi'}$ vanishes is topologically $D^2$,  this denotes a two-dimensional disc-shaped bubble.

The Killing vector $v_H :=\partial_t +\omega_\psi \partial_\psi+ \omega_\phi \partial_\phi $ becomes null   at $y=-1/\nu\ (-1\le x  \le 1)$ and timelike at infinity, and so $y=-1/\nu\ (-1\le x  \le 1)$  corresponds to a Killing horizon with the surface gravity
\begin{align}
\kappa = \frac{(1-a^2)^{3/2} (1-\gamma )^2 \sqrt{\nu  (\nu   +1)(\gamma +\nu)^{-1}}}{c_3 \ell  \left(c^3 \left(1-\nu-a c_1 \right)+s^3 (b-a) (\gamma -\nu   )\right)},
\end{align}
where the horizon angular velocities $\omega_\psi,\omega_\phi$ are given by
\begin{align}
 \omega_\psi=-\frac{c_3 \omega_\phi}{d_3}=\frac{v_0 c_3 \kappa }{2 (1-\gamma ) \sqrt{\nu  (\nu +1)(\gamma +\nu )(1-a^2)}}.\hspace{-0.2cm}
 \end{align}
The two Killing vectors, $v_{\phi}:=\partial_{\phi}$ and $v_{\phi'}:=\partial_{\phi'}$, vanish on  $x=-1$ and $x=1$, respectively, following ${\rm det}(v_\phi,v_{\phi'})=-1$. Hence the theorem in Ref.~\cite{Hollands:2007aj} shows that the spacial cross section of  the Killing horizon at $y=-1/\nu$ is topologically $S^3$.
\vspace{0.05cm}
\paragraph*{Absence of curvature singularities and CTCs: }
The metric and its inverse seem divergent at the surfaces $H(x,y)=0$ and $D=0$, where the curvature singularities may appear, e.g.,  the Kretschmann scalar can be written as $R^{\mu\nu\rho\lambda}R_{\mu\nu\rho\lambda} \propto H(x,y)^{-6} D^{-6}$.
 It is easy to check that the five terms in eq.~(\ref{eq:H2}) are nonnegative within the range specified in eq.~(\ref{eq:a,b}) for eq.~(\ref{eq:Bd-sol}) and $-1\leq x \leq 1$, $-1/\nu \leq y \leq -1$, hence $H(x,y)\ge 0$. 
Therefore, $H(x,y)=0$ if and only if all of five terms vanish at the same time, however, this does not occur.  
Thus,  it turns out that $H(x,y)>0$ everywhere outside and on the horizon.
The positivity of $H(x,y)$ directly implies $D>0$ because $H(x,y) D  > c^2 [H(x,y)-H(y,x)] >0$, where $H(x,y)-H(y,x)$ can be written as the sum of nonnegative terms which do not vanish at the same time.

Having shown the positivity of $H(x,y)$ and $D$, one can also show the absence of curvature singularities at the boundaries of the $C$-metric coordinate system $(x,y)$. 
(i)One can see that $y=-1/\nu$ is a regular horizon by the  transformation $dx^i \to dx^i -v_H^i (1-\nu^2)dy/(2\kappa\nu G(y))$ and the gauge transformation $A \to A -d[(1-\nu^2)(A_i|_H v_H^i)\int^y dy/(2\kappa\nu G(y))]$  ($i = t,\psi,\phi$) where $A_i|_H:=A_i(y=-1/\nu)$. 
(ii) Near $x=\pm 1$, in terms of  $x=\pm 1\mp C_\pm r^2$ with positive constants $C_\pm$, the metric behaves as $ds^2 \simeq \alpha_{00} dt^2 +2\alpha_{01} dt d\phi_1 + \alpha_{11} d\phi_1^2 + \beta_1 dy^2 + \beta_2 (dr^2 + r^2 d\phi_2^2)$, 
where $\alpha_{ij}(y)\ (i,j=0,1)$ is a certain regular, nondegenerate matrix and $\beta_i(y)$ $(i=1,2)$ are some positive function. We have set $(\phi_1,\phi_2) = (\psi,\phi)$ at $x=-1$ and $(\phi_1,\phi_2)=(\psi-\phi,\phi)$ at $x=1$. 
(iii) It can be also shown that near $y=-1$, in terms of $y=-1-C_0 r^2$ with a positive constant $C_0$ and $(\phi_1,\phi_2)=(\phi,\psi)$, the metric takes a similar regular form, where the role of $y$ is replaced with $x$.
(iv) Lastly, one can show that at the point $(x,y)=(1,-1)$, the metric is regular because by introducing the new coordinates $(t',\psi',\phi',x,y):=(C_1 t, \psi - \phi, \phi, 1-C_2(1+\nu) r^2 \sin^2\theta ,-1-C_2(1-\nu)r^2 \cos^2\theta)$ with appropriate constants, $C_1,C_2$, the metric behaves as the Minkowski metric $ds^2 \simeq -dt'^2 + dr^2 + r^2 (\cos^2\theta d\psi'^2+\sin^2\theta d\phi'^2 )d\theta^2$.
Therefore, the spacetime does not have any curvature singularities outside and on the horizon.
Moreover, the absence of CTCs requires that the two-dimensional angular part $g_{IJ}\ (I,J=\phi,\psi)$ of the metric is nonnegative on and outside the horizon (which is determined from the non-negativity of the trace and determinant for $-1<x<1$ and $-1/\nu<y<-1$).  We can show the non-negativity numerically, and hence the absence of CTCs in the DOC.

\paragraph*{Physical quantities: }
The ADM mass, two angular momenta are computed as
\begin{align}
 &M=
 \frac{3 \pi  \ell^2 (1+2 s^2) (\gamma +\nu ) (c_3
   (1-\nu )-d_1)}{4 G_5 \left(1-a^2\right) (1-\gamma )^2 (1+\nu)},\label{eq:MQ}\\
   &  J_\psi = c^3 J_1 + s^3 J_2,\quad J_\phi = c^3 J_2 + s^3 J_1,
\end{align}
where $J_1$ and $J_2$ correspond to the angular momenta in the neutral case, 
\begin{align}
   &J_1=\frac{\pi  \ell^3 v_0 \left(c_3 d_2-c_1   c_2 c_3-(c_3-b c_2) d_1\right)}{4G_5 \nu \left(1-a^2\right) (1-\gamma )^3  },\\
&J_2 =\frac{\pi  \ell^3 v_0 (a-b) \left(2 c_3 \nu +d_1\right)}{2G_5
   \left(1-a^2\right) (1-\gamma ) (1+\nu )}.
\end{align}

\begin{widetext}
\begin{figure}[t] 
\begin{center}
\begin{minipage}[t]{2.1\columnwidth}
  \includegraphics[width=5.8cm]{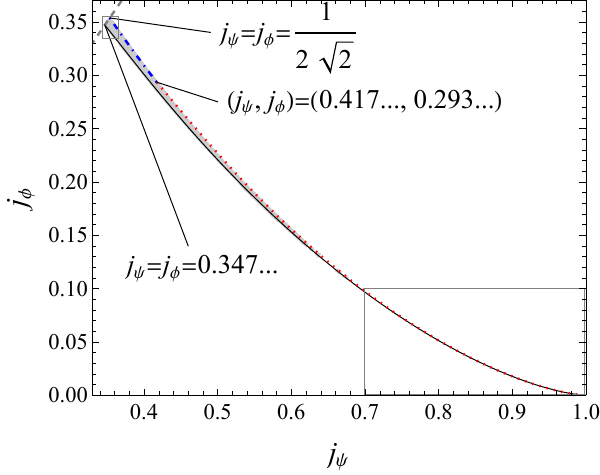}
     \includegraphics[width=5.8cm]{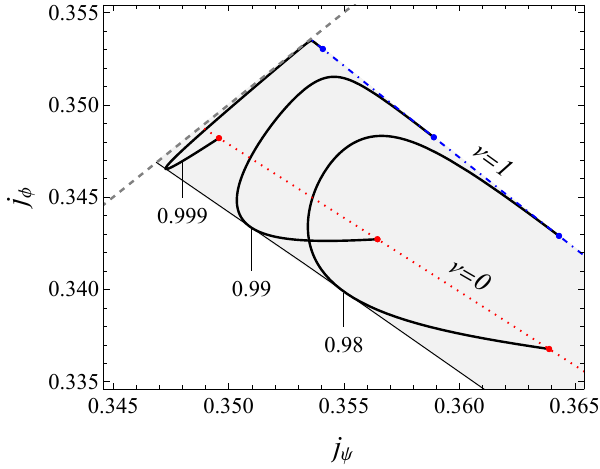} 
  \includegraphics[width=5.8cm]{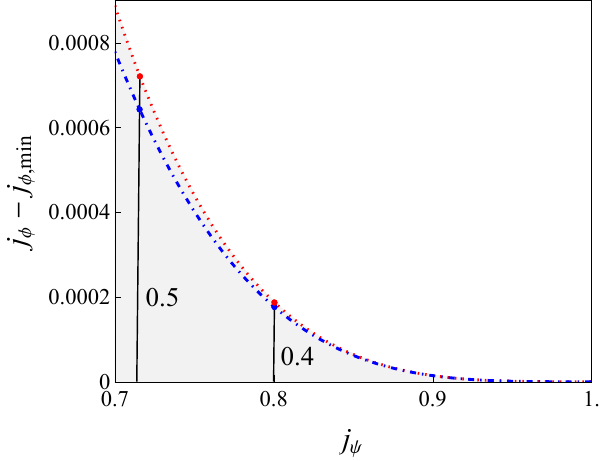} 
  \caption{The allowed region of two angular momenta and the constant curves for an electric charge ($\tanh \alpha$).  
In the left panel, the whole region is shown by the shaded region. 
The middle and right panels display the close-ups of  $0.348\ldots \le j_\psi\le 0.365$ and  $0.7\le j_\psi\le 1$, respectively, in the left panel. 
The red dashed and blue dotted-dashed lines represent $\nu=0$ and $\nu=1$, respectively, 
which 
have an intersection at the point $(j_\psi,j_\phi)= (0.417\dots,0.293\dots)$.
In the middle panel, some $\tanh\alpha={\rm const.}$ curves are plotted near the BPS limit ($\tanh \alpha\to\infty$), 
where the curve approaches the black dashed line $j_\psi=j_\phi$. 
As seen in the right panel, for $j_\psi \to 1\ (\alpha \to 0)$, the region gets narrowed toward a point $(j_\psi,j_\phi)=(1,0)$, where $j_\psi$ and 
the difference $j_\phi$-$j_{\phi,{\rm min}}$ from the lower bound of $j_\phi$ are plotted due to lacking in resolution. 
\label{fig:jjplot}}
     \end{minipage}
     \end{center}
\end{figure}
\end{widetext}

The electric charge over a three-dimensional closed surface $S$ surrounding a horizon and a bubble  which is  defined as
 \begin{align}
Q := \fr{8\pi G_5} \int_{S} \left(\star F+\frac{1}{\sqrt{3}}F\wedge A \right) = - \frac{2M\tanh(2\alpha)}{\sqrt{3}},
\end{align}
which obviously
follows the BPS bound $M \geq \frac{\sqrt{3}}{2} |Q|$.

In addition to these conserved charges, the existence of the disc-shaped bubble $D$ at $x=1$  enables one to define the magnetic flux (this is not  a conserved charge) as
 \begin{align}
  q := \fr{4\pi} \int_D F 
  =\frac{\sqrt{3} \ell s c d_1  v_0 (b d_2 s (\gamma -\nu )-2 c \nu 
   \tilde{d}_2)}{2
   (c^2 d_2^2 (\gamma -\nu )+2 \nu  s^2 {\tilde{d}_2}^2)},
\end{align}
where $\tilde{d}_2:=d_2-(\gamma -1) (\nu
   -1) (\gamma +\nu)$.
The area of a constant time slice through the horizon is written as
\begin{align}
&A_H = 8 \ell^3 \pi ^2 c_3 \sqrt{\nu(\gamma +\nu)(1+\nu)^{-1}(1-a^2)^{-3} }   (1-\gamma )^{-2}\nonum
& \qquad  \times \left(c^3   \left(1-\nu -a c_1\right)-s^3 (a-b) (\gamma -\nu )\right).
\end{align}

\paragraph*{Phase space:}
Now, to consider the phase diagram,  we introduce the angular momenta and horizon area normalized by the  mass scale $r_M :=\sqrt{8 G_5 M/3\pi}$  as $j_\psi := 4G_5 J_\psi/\pi r_M^3,\ j_\phi := 4G_5J_\phi/\pi r_M^3$ and $a_H := \sqrt{2}A_H/\pi^2 r_M^3$.
Similarly, 
we introduce the normalized angular momenta  $j_\psi^{\rm CY}$, $j_\phi^{\rm CY}$ and the normalized  horizon area $a_H^{\rm CY}$ for the Cveti\v{c}-Youm solution~\cite{Cvetic:1996kv}. 
Figure~\ref{fig:jjplot}  shows the allowed region of regular black holes without CTCs  in the $j_\psi$-$j_\phi$ plane. 
The angular momenta have the ranges $0<j_\phi< 1/(2\sqrt{2})=0.354\ldots, \ 0.347\ldots<j_\psi<1$.

\begin{widetext}
\begin{figure}
\begin{center}
\begin{minipage}[tb]{2\columnwidth}
\includegraphics[width=7cm]{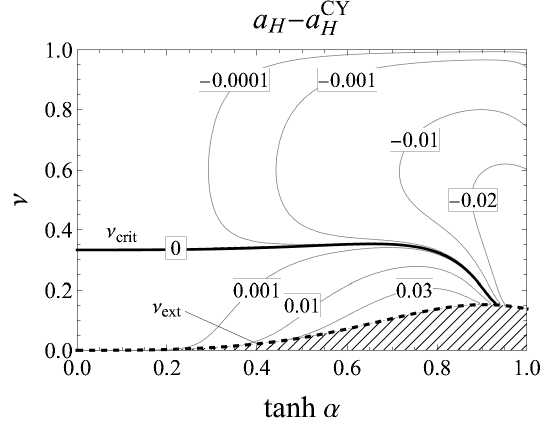} 
  \includegraphics[width=7cm]{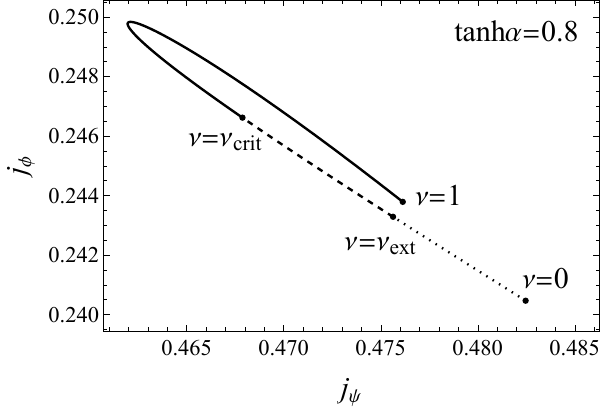} 
\caption{The comparison  of the dimensionless horizon areas with the Cveti\v{c}-Youm black hole for the same conserved charges $(M,J_\psi,J_\phi,Q)$.  
The left panel displays the contour plots of $a_H-a_H^{\rm CY}$ in the $\tanh\alpha$-$\nu$ plane. For $\nu_{\rm ext}<\nu<1$ (the white region),  a Cveti\v{c}-Youm black hole with the same charges exists for each $(\tanh\alpha,\nu)$, while for $0<\nu<\nu_{\rm ext}$ (the hatched region), there is no Cveti\v{c}-Youm black hole with the same charges.
The right panel displays the curve corresponding to $\tanh\alpha=0.8$ plotted in the $j_\psi$-$j_\phi$ plane as an example. 
For $\tanh\alpha <0.940... $, the capped black hole with  $\nu_{\rm ext}<\nu<\nu_{\rm crit}$ ($\nu_{\rm crit}<\nu<1$) has larger (smaller) entropy than the Cveti\v{c}-Youm black hole, whereas for $0.940... <\tanh\alpha<1$, the former always has smaller entropy than the latter.
\label{fig:entropy}}
\end{minipage}
\end{center}
\end{figure}
\end{widetext}

The solid curves correspond to various $Q={\rm const.}$ (various values of $\tanh \alpha$), each of which
have end points at $\nu=0$ and $\nu=1$.
The region is bounded by the dashed line  $j_\psi=j_\phi$, which can only be reached at the BPS limit $M=\sqrt{3}|Q|/2\ (\alpha\to \infty)$. Since the metric is no longer asymptotically flat at the BPS limit,
the capped black hole does not admit equal angular momenta $j_\psi=j_\phi$.
Moreover, since each $Q={\rm const.}$ curve, 
as seen in the middle panel of Fig.~\ref{fig:jjplot},
 is not closed, the  capped black hole is uniquely specified by the conserved charges of its mass, two angular momenta and electric charge. 
As in the right panel of Fig.~\ref{fig:entropy}, each $Q={\rm const.}$ curve can be decomposed into three parts, (i) $0<\nu< \nu_{\rm ext}$, (ii) $\nu_{\rm ext}<\nu< \nu_{\rm crit}$ and (iii) $\nu_{\rm crit}<\nu< 1$,
where at $\nu=\nu_{\rm ext}$, the Cveti\v{c}-Youm black hole with the same conserved charges becomes extremal and at $\nu=\nu_{\rm crit}$ has the same entropy as the capped black hole.
Hence, for (i), 
the capped black hole has larger angular momenta beyond the extremal Cveti\v{c}-Youm black hole,
whereas for (ii) and (iii), the capped black hole and the Cveti\v{c}-Youm black hole have the same mass, two angular momenta and electric charge, which obviously shows non-uniqueness  property of black holes even under the assumption that the horizon cross section has the topology of $S^3$.
In particular,  for (ii),  the capped black hole has larger entropy than the Cveti\v{c}-Youm black hole, and hence 
the capped black hole with a large bubble becomes thermodynamically more stable.
As seen from the left panel of Fig.~\ref{fig:entropy}, this regime disappears for $\tanh\alpha > 0.940\ldots$. This also implies that too much amount of the electric charge destabilizes the bubble structure.

It is worth noting that, due to the disc-shaped bubble, there would be more general solutions with five independent parameters of the mass, two angular momenta, electric charge and magnetic flux~\cite{Kunduri:2013vka}.

\vspace{-0.3cm}
\section*{Acknowledgement}
\vspace{-0.3cm}
R.S. was supported by JSPS KAKENHI Grant Number JP18K13541. 
S.T. was supported by JSPS KAKENHI Grant Number 21K03560.

\end{document}